\documentclass[conference]{IEEEtran}
\IEEEoverridecommandlockouts
\usepackage{cite}
\usepackage{amsmath,amssymb,amsfonts}
\usepackage{algorithmic}
\usepackage{graphicx}
\usepackage{textcomp}
\usepackage{xcolor}
\def\BibTeX{{\rm B\kern-.05em{\sc i\kern-.025em b}\kern-.08em
    T\kern-.1667em\lower.7ex\hbox{E}\kern-.125emX}}
\begin{document}

\title{Checkpoint/Restart for Lagrangian particle mesh with AMR in community code FLASH-X. \\}
\author{\IEEEauthorblockN{Rajeev Jain}
\IEEEauthorblockA{\textit{Mathematics and Computer Science Division } \\
\textit{Argonne National Laboratory}\\
Lemont, IL USA \\
jain@anl.gov}
\and
\IEEEauthorblockN{Klaus Weide}
\IEEEauthorblockA{\textit{Department of Computer Science} \\
\textit{The University of Chicago}\\
Chicago, IL USA \\
kweide@uchicago.edu
}
\and
\IEEEauthorblockN{Saurabh Chawdhary}
\IEEEauthorblockA{\textit{} \\
chaw0023@umn.edu}
\and
\IEEEauthorblockN{Thomas Klostermann}
\IEEEauthorblockA{\textit{Mathematics and Computer Science Division } \\
\textit{Argonne National Laboratory}\\
Lemont, IL USA \\
tklosterman@anl.gov}
}

\maketitle

\begin{abstract}
In this work we present the design decisions and advantages for accomplishing cross mesh format checkpoint-restart in community code FLASH-X. AMReX and Paramesh are the two AMR mesh formats developed and supported by FLASH-X. We also highlight strong and weak scaling study of existing HDF5 I/O checkpoint writing along with new ideas and results (presented during talk) for utilizing heterogeneous compute architectures for improved I/O performance.
\end{abstract}


\section{Background}
FLASH is a modular, parallel multiphysics simulation code capable of handling general compressible flow problems found in many astrophysical environments. It has been used in production for several decades in many areas of scientific research \cite{Dubey2009}. FLASH-X is a new version of FLASH targeted towards heterogeneous architecture and designed to incorporate performance portability. The current version is capable of nearly ideal weak-scaling up to 10,000 MPI ranks on the Summit computer at the Oak Ridge Leadership Computing Facility (OLCF).
FLASH-X uses HDF5 parallel I/O libraries in both serial and parallel forms, where the serial version collects data to one processor before writing it, while the parallel version has every processor writing its data to the same file. It should be noted that the checkpointing studied here is application controlled and not transparent checkpointing. The checkpoint files in FLASH-X are used for restarting a simulation after it was interrupted due to planned or unplanned events. 
The checkpoint files contain all the information required for restarting the simulation. The data can be saved in full precision of the code (8-byte reals) and includes all of the variables, species, grid construction data, scalar values, as well as meta-data about the run. In many applications of FLASH-X, checkpoints can also serve as the primary output result of the simulation. Various visualization tools support display and analysis of data stored in FLASH-X's checkpoint files.
\section{Cross-Checkpoint/Restart in FLASH-X}
Checkpointing and restart functionality has existed in FLASH for more than a decade, provided by the AMR package Paramesh. However, the recent addition of a second AMR option (implementation details available in \cite{jared2018}), AMReX \cite{AMReX}, has spurred development of a technique to restart using a different mesh package than the mesh package that created the initial simulation checkpoints. This procedure is here called a cross-checkpoint-restart. Multiple grid implementations and support for cross-checkpoint-restart increase our confidence
in the validity of simulation results and their independence from the underlying grid representation. The refinement schemes, mesh storage, and optimizations provided by both the grid formats (i.e. Paramesh and AMReX) are different and may benefit the total run-time of a simulation at different stages in the evolution. This makes the case for cross-checkpointing across mesh types. Two cases are studied and implemented, (i) an initial simulation using AMReX writes checkpoint files followed by restart using Paramesh, and (ii) an initial simulation using Paramesh writes checkpoint followed by restart using AMReX. 
AMReX provides a native I/O format for checkpointing that can be used to restart an AMReX simulation but this does not allow for reading of the AMReX-checkpoint files by native FLASH-X tools or, in other words, for cross-checkpoint-restart. In both cases of cross-checkpoint-restart, we opted to use FLASH-X's checkpoint HDF5 file format irrespective of the mesh format. For the first case, the AMReX mesh is converted to a Paramesh-like representation in the I/O unit of FLASH-X, immediately before writing the results to a native FLASH-X checkpoint file. This also provides a way to map the solution variables onto the FLASH-X checkpoint format. For the second case, the Paramesh checkpoint file is loaded into FLASH-X's native I/O unit; then an AMReX-format mesh is manually created and passed to AMReX during initialization.
A similar design is used for reading checkpoint data on particles (Lagrangian mesh) from one format to the other. In AMReX, particles are identified by a combination of cpu id (rank of birth processor of a particle) and tag, whereas for Paramesh particles are identified by a unique tag number assigned each particles. New changes were introduced to make the two formats identical and generate cpu id for Paramesh grid. This new capability can help scientists utilize the power of both native Paramesh grid and externally developed AMReX grid.

The two grid implementations Paramesh and AMReX result in different refinement histories, and the simulation results obtained from one format is not expected to be the same as obtained from another. In order to compare and verify the results from cross-checkpointing with different meshes, we exploit "sfocu", a tool provided by FLASH-X to compare two different checkpoint files. It is used to benchmark results in continuous integration testing. This tool, sfocu reports, mag error, among other comparison metrics for variables defined on mesh. Mag error(\textit{mE}) for simulation variable appearing in two checkpoint files as \textit{a} and \textit{b} is computed as follows:
\begin{equation}
\label{eq:magerror}
mE = \frac{sup \left|a-b  \right|}{max[(sup \left|a\right|), (sup\left| b \right|), 10^{-99}]}
\end{equation}

\begin{figure}
\centerline{\includegraphics[width=90mm,scale=1.5]{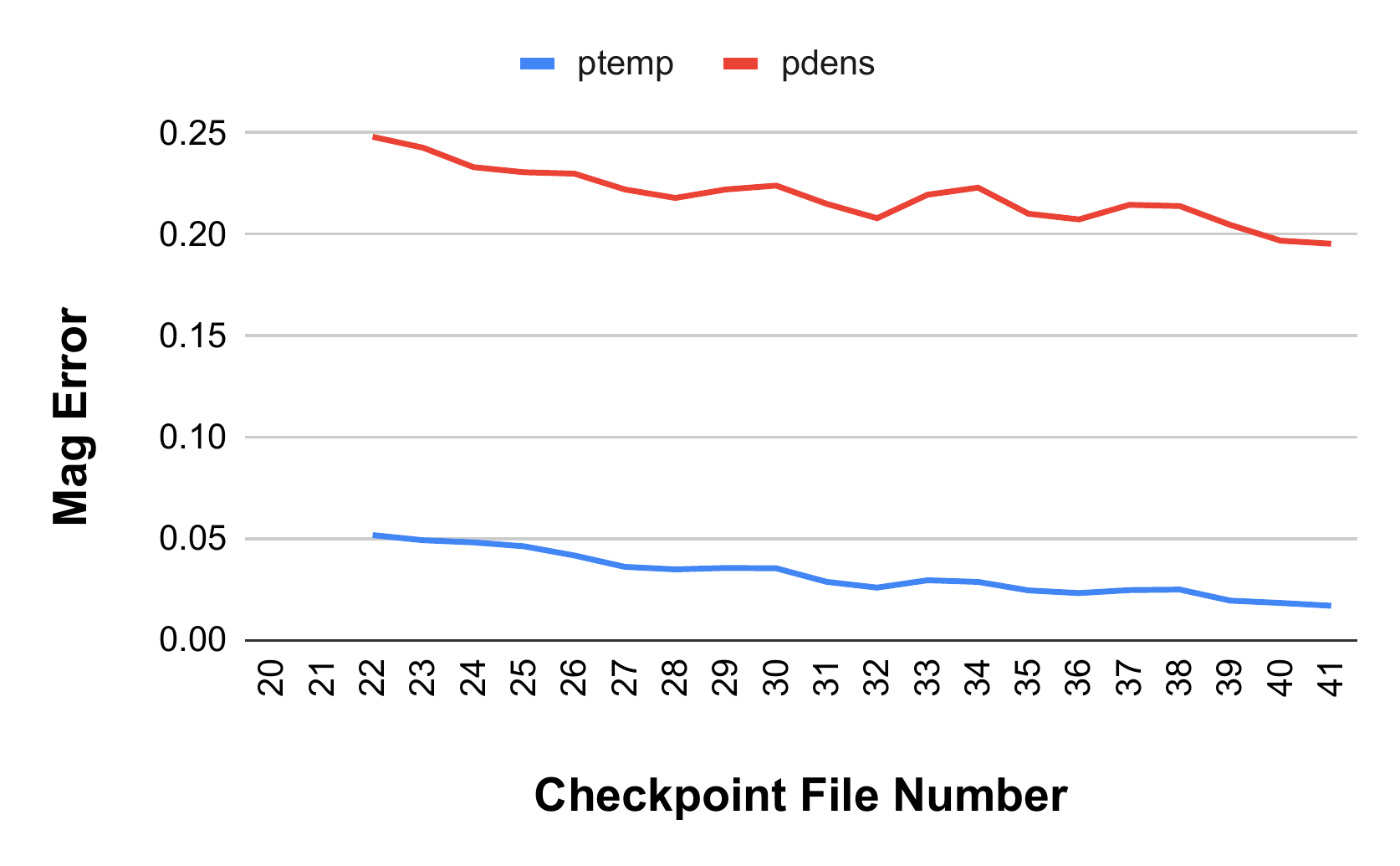}}
\caption{Relative magnitude error as defined by equation \ref{eq:magerror} for Lagrangian (particle) variable temperature - "ptemp" and density - "pdens" for Paramesh and Paramesh restart from AMReX checkpoint file}
\label{lag_err}
\end{figure}

\begin{figure}
\centerline{\includegraphics[width=90mm,scale=1.5]{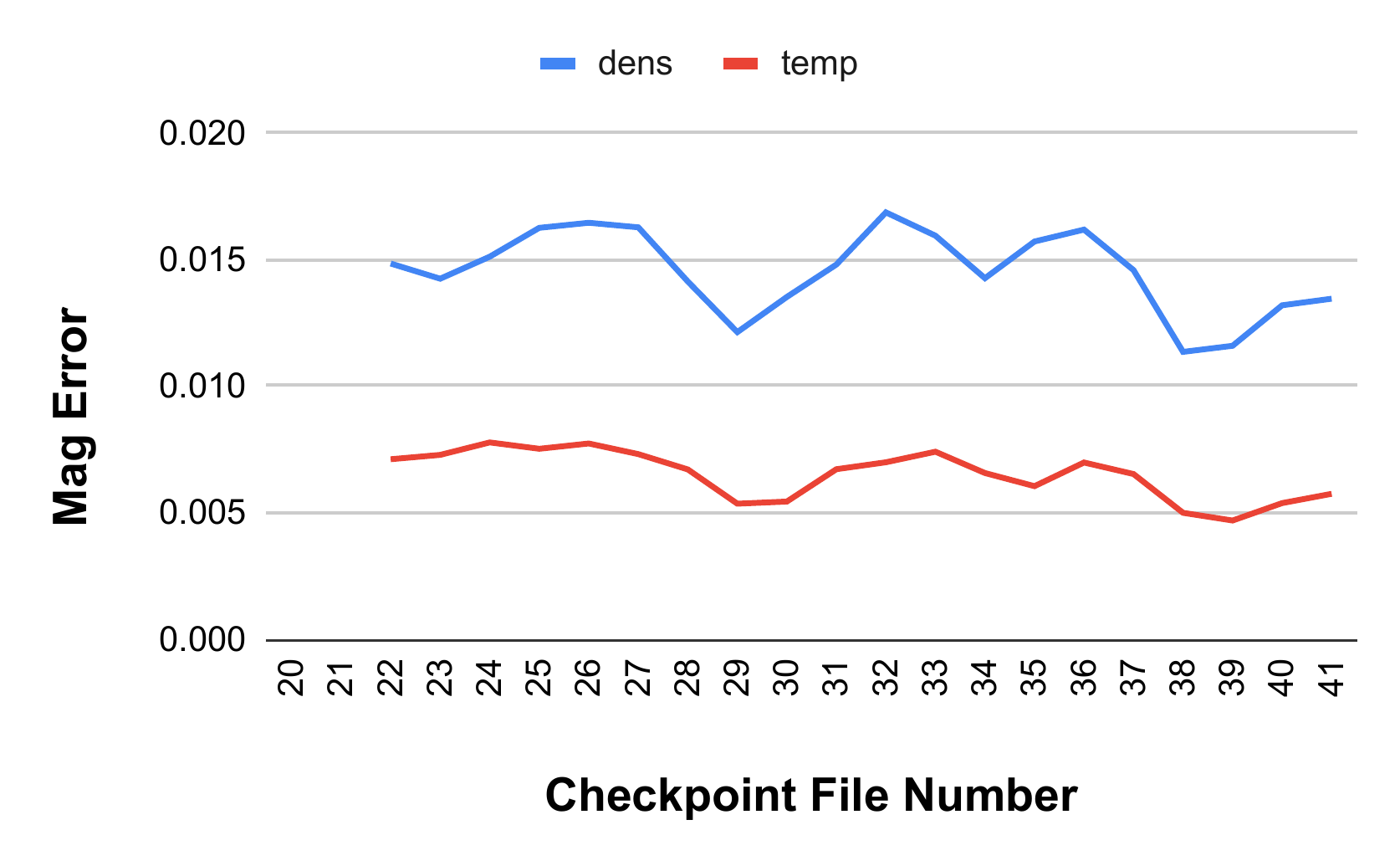}}
\caption{Relative magnitude error as defined by equation \ref{eq:magerror} for Eulerian variable temperature - "temp" and density - "dens" for Paramesh and a Paramesh restart from AMReX checkpoint file}
\label{eul_err}
\end{figure}

Figures \ref{lag_err} and \ref{eul_err} show \textit{mE} for a set of checkpoint files, numbered 22 to 41 in both figures, For, the two compared simulations, one is run using paramesh until checkpoint 21 then restarted using this checkpoint on AMReX mesh from checkpoints 21-41; the second is similar but run with AMReX till checkpoint 21 then restarted on paramesh mesh to generate checkpoint files from 21 through 41.Figure \ref{lag_err} plots \textit{mE} for these simulations but for the Lagrangian particle variables, temperature and density, for checkpoint number 22 to 41. It can be seen that \textit{mE} decreases gradually as the simulation progresses. We should not expect the two results to be numerically equal or start out close to each other as the underlying AMR grid is different. Moreover, the two mesh packages have different refinement strategies resulting in potentially diverging mesh pattern. The plots here indicate that even though \textit{mE} is non-negligible, it is diminishing as the simulation progresses. Figure \ref{eul_err} highlights a much lower \textit{mE} for Eulerian variables compared to Lagrangian variables. This could be because the same particle locations may be mapped to different Eulerian mesh cells and the Lagrangian variables are calculated after interpolation from new Eulerian mesh with potentially different values.

\section{Checkpointing I/O Performance}
\subsection{Strong and weak scaling of synchronous I/O}
\begin{figure}
\centerline{\includegraphics[width=90mm,scale=1.55]{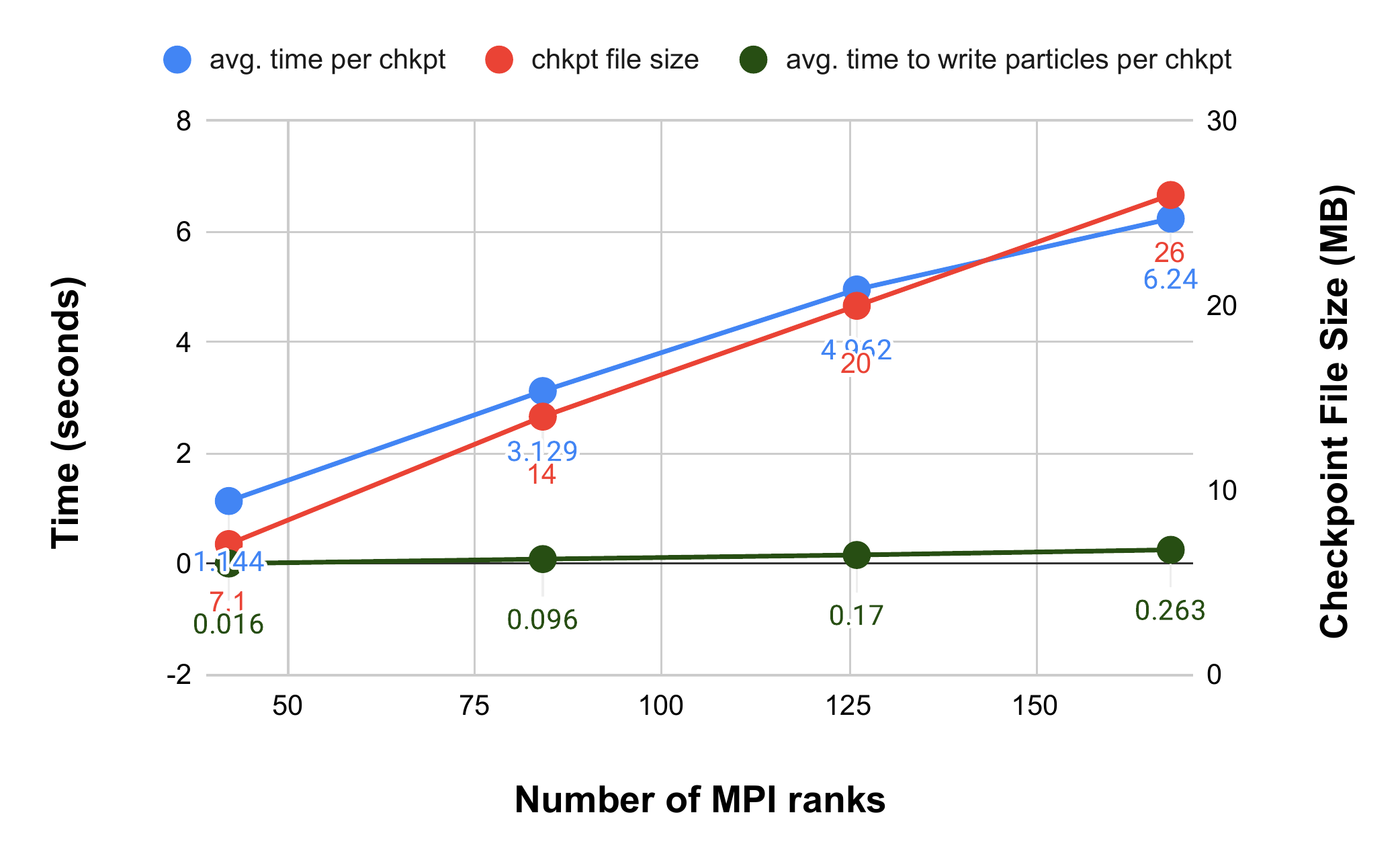}}
\caption{Weak scaling: Variable problems size with constant blocks/proc, run on up to 184 MPI ranks}
\label{weaksc}
\end{figure}

This section presents I/O performance results on a relatively small number of MPI ranks (up to 184) using the 2D Sedov problem (a compressible flow explosion problem widely used for verification of shock-capturing simulation codes) with tracer particles in FLASH-X. A weak and strong scalability study for I/O writing from the AMReX grid with HDF5 checkpoint files is performed. Results are obtained using a GNU compiler (gcc 6.4.0) and HDF5 Version 1.10.2 on Summit\cite{summit}. Figure \ref{weaksc} shows particles writing time, checkpoint writing time and checkpoint file size. For all the runs, the number of blocks written per processor is set at 18. It can be seen that size of checkpoint file increases from 7.1MB (number of particles is 8x8 and number of blocks is 16x16, 1 node run) to 26MB (number of particles is 20x20 and number of blocks is 40x40, 4 node run).  Figure \ref{weaksc} indicates that the time to write particles and checkpoint file increases almost linearly with an increase in the number of processors. The per processor writing load is same on all processors for all the runs. The increasing trend in write times is attributed to the small size of data written and high communication cost by HDF5 file writing routines.

\begin{figure}
\centerline{\includegraphics[width=90mm,scale=1.55]{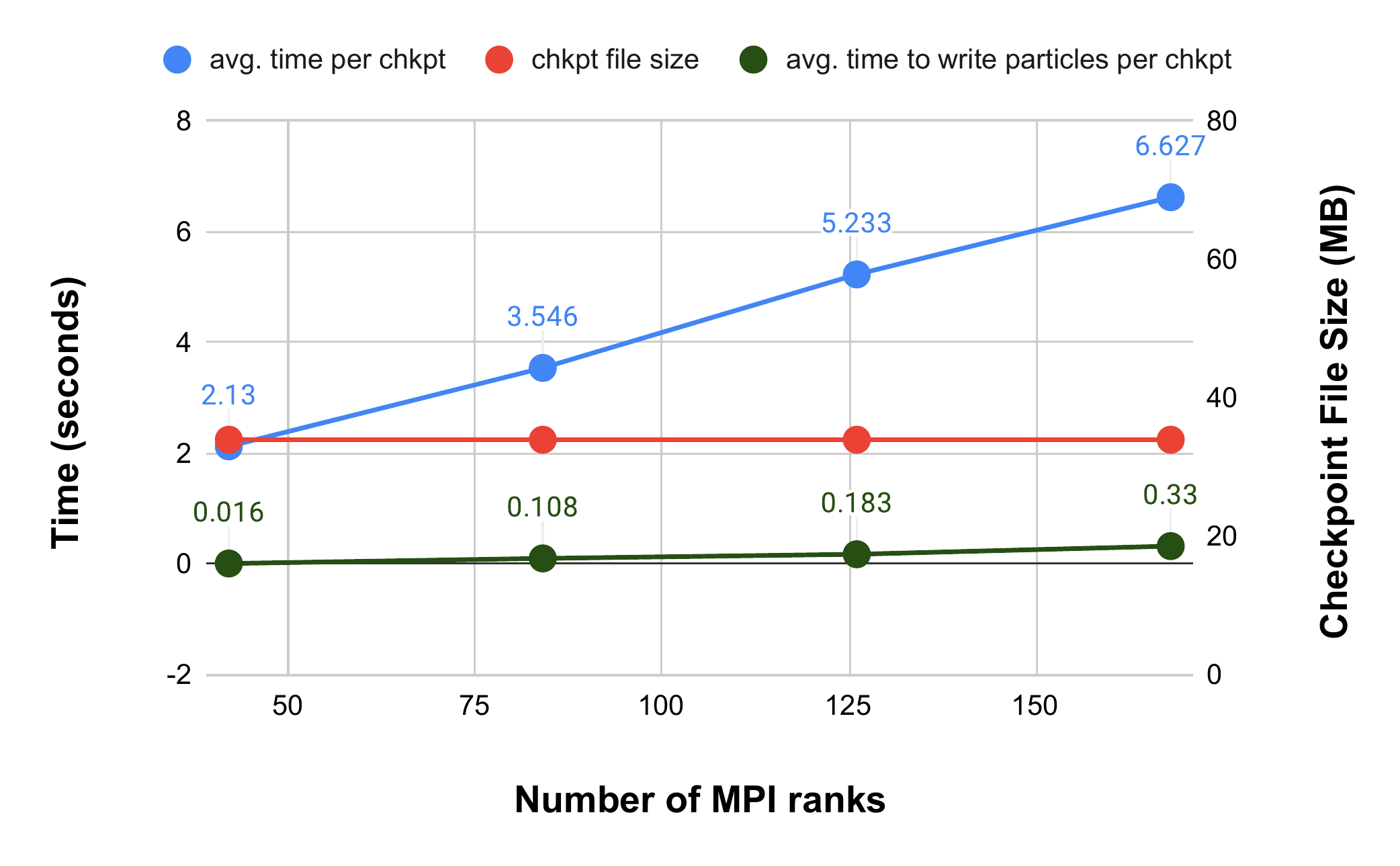}}
\caption{Strong scaling: Running the same problem on an increasing number of nodes, with 42 ranks/node}
\label{strongsc}
\end{figure}

Figure \ref{strongsc} shows the result of increasing the the number of ranks while running the same problem size. In each run, a 34MB checkpoint file is written to disk by different number of processors. We report here the average time to write (i) particles only, and (ii) overall checkpoint file of the 2D Sedov simulation with 24x24 particles and 46x46 blocks. With a four times increase in the number of MPI ranks, the checkpoint files also takes four times as long to write. This is due to increased communication costs with higher number of ranks (there are fewer blocks per rank). The I/O routines currently use HDF5's \textit{collective mode} with hyperslabs without any GPU utilization, these routines can be further optimized to improve the time for writing a single checkpoint file.
\subsection{Asynchronous I/O}
Strong and weak scalability results, although on relatively small number of processors, are clear motivations for testing recent new developments in HDF5's Virtual Object Layer (VOL) \cite{byna2020exahdf5} that allows different underlying storage mechanisms. The authors in that paper report that using asynchronous (asynch) I/O feature of VOL connector the total I/O time required by an application was reduced by up to 4 times. Here, we use Argobots \cite{argobots}, a low level light weight threading library as the VOL. Using asynchronous communication pattern enables the application thread to believe that the checkpoint file is written and it can continue with the simulation. In reality, a copy of the data is made (caching and other memory optimal schemes are also available) and given to the threads for writing, while the main application continues with the simulation. This enables asynchronous I/O to achieve the goal of completely overlapping computation with communication at the cost of memory. As mention in \cite{byna2020exahdf5},an event set (ES) id is created by HDF5 to manage the asynchronous operations. This ES id is created before we begin any I/O operation, and is passed as an argument to all async I/O functions. At the time of writing the last checkpoint, ES "wait" is called to wait for any unfinished I/O operations. I/O cannot be 100\% overlapped with computation as no further computation may be required when the simulation is ending.

Table \ref{asynctable} shows our preliminary results obtained by running FLASH-X's 2D Sedov test problem. \textit{Avg/proc} is the average time taken by an MPI rank to write on checkpoint file. \textit{Num calls} is the number of checkpoint files written by the simulation, in this test problem 8 checkpoint files are written. The last checkpoint file cannot be overlapped with computation as it is written after all the computation is done. \textit{I/O time} is total time spent on for writing checkpoint files and finally percentage of simulation time that was spent on I/O operation is shown. On 6 MPI ranks, we run simulations with both asynchronous and synchronous versions of I/O routines. It is found that asynchronous I/O deliver savings in compute time which is significant not only for I/O but also in terms of overall simulation time. A savings of 4.5\% of total simulation time is seen for this particular test case. Readers should note that at present "async" support in FLASH-X and HDF5 routines are in development phase and the results presented here are experimental in nature.
\begin{table} [!ht]
\centering
\begin{tabular}{ || c c c c c || }
\hline
I/O & Avg/proc(s) & Num calls & I/O Time(s) & \% of Tot. on I/O  \\ [0.5 ex]
\hline \hline
Sync & 9.876 & 8 & 11.25 & 5.518 \\
Async & 0.816 & 8 & 2.12 & 1.069 \\
\hline
\end{tabular}
\caption{Comparison of regular synchronous I/O with latest asynchronous version of I/O}
\label {asynctable}
\end{table}


\section{Discussion and Conclusion}
FLASH-X has robust checkpoint and restart mechanisms, here we highlight the new key design decision for enabling cross-checkpoint-restart (switching mesh types while restarting simulations). This feature will be required for obtaining performance/accuracy gains at different stages of the simulation with different underlying grid optimization available with underlying AMR grid representation. Such checkpoint-restart functionality will also help increase our confidence in the simulation results in terms of their independence of underlying grid representation. 
We also present strong and weak scalability results for writing a checkpoint file in FLASH-X. Results indicate that there is clear scope for improvement in I/O performance. We highlight the ongoing work with the use of new VOL asynchronous I/O plugin developed in HDF5 that has potential for substantial I/O speedup. 
We also acknowledge that the present data may be based on limited set of test cases and relatively small number of MPI ranks. We wish to address these concerns in the future.
\section*{Acknowledgment}
    We thank Dr. Houjun Tang at Berkeley Lab for help with Asynchronous HDF5. FLASH was developed, in part, by the DOE NNSA ASC- and DOE Office of Science ASCR-supported Flash Center for Computational Science at the University of Chicago. Funding for FLASH-X and ExaStar was supported by the Exascale Computing Project (17-SC-20-SC), a collaborative effort of the U.S. Department of Energy Office of Science and the National Nuclear Security Administration.
This research used resources of the Oak Ridge Leadership Computing Facility at the Oak Ridge National Laboratory, which is supported by the Office of Science of the U.S. Department of Energy under Contract No. DE-AC05-00OR22725.

\bibliographystyle{IEEEtrans}
\bibliography{ecp_astro}

\end{document}